\documentclass[12pt]{article}

\usepackage[utf8]{inputenc}
\usepackage[T1]{fontenc}
\usepackage{lmodern}
\usepackage{setspace}
\usepackage[margin=1in]{geometry}
\usepackage{amsmath,amssymb,amsfonts}
\usepackage{booktabs}
\usepackage{natbib}
\usepackage{hyperref}
\usepackage{enumitem}
\usepackage{graphicx}

\hypersetup{
    colorlinks=true,
    linkcolor=blue,
    citecolor=blue,
    urlcolor=blue
}

\title{\textbf{When Do Markets Fully Process Public Information? Evidence from Real-Time Prediction Markets}}

\author{
Giovanni Angelini\thanks{Department of Economics, University of Bologna, g.angelini@unibo.it.}
\and
Luca De Angelis\thanks{Department of Economics, University of Bologna, l.deangelis@unibo.it.}
}

\date{\today}

\begin{document}

\maketitle

\singlespacing

\begin{abstract}
How efficiently do markets update beliefs when public information arrives in rapid sequence? We use a real-time prediction market setting that combines binary payoffs, precisely observed public signals, and high-frequency market data, allowing us to compare market price changes with changes in a benchmark probability implied by publicly available information. We first show that prices are informative and become more accurate as resolution approaches. During the event, prices respond rapidly to public signals and move in the expected direction. However, directional responsiveness is not the same as efficient updating. Relative to an out-of-sample benchmark probability model, a one-minute change in the benchmark probability is associated with only about a 0.64-for-one contemporaneous change in market prices. The missing adjustment predicts future price drift over the following several minutes, including drift net of subsequent changes in the benchmark probability. We then study the mechanisms underlying this gradual adjustment. Salient public signals are incorporated relatively quickly in liquid markets, but the same signals generate substantially greater underreaction when liquidity is low. Underreaction gaps associated with salient states also predict stronger subsequent drift. The evidence therefore points to gradual price discovery shaped by the interaction between attention and trading frictions. The results contribute to the literatures on prediction markets, market efficiency, and behavioral finance. More broadly, they show that markets can aggregate public information quickly without necessarily incorporating it fully on impact. Market-implied probabilities are often directionally correct, yet adjustment remains incomplete and predictably depends on liquidity and salience.
\end{abstract}


\onehalfspacing

\section{Introduction}

When do markets fully process public information? This question is central to economics because prices are often used as sufficient statistics for beliefs, expectations, and decision-relevant information. In the efficient-markets benchmark, public news should be incorporated rapidly and without predictable subsequent price movements \citep{Fama1970}. Yet this benchmark need not hold when attention, information processing, and trading are costly \citep{GrossmanStiglitz1980}. Models of investor sentiment, overconfidence, representativeness, and gradual information diffusion provide mechanisms for both underreaction and overreaction to news \citep{BarberisShleiferVishny1998,DanielHirshleiferSubrahmanyam1998,HongStein1999}, while limited attention can delay price adjustment even when information is public \citep{DellaVignaPollet2009,HirshleiferLimTeoh2009}. A persistent empirical difficulty, however, is that the econometrician rarely observes both the precise arrival of information and the asset's relevant fundamental value. This paper studies real-time belief updating in a setting where these objects can be measured unusually well. We use live National Basketball Association (NBA) event contracts traded on Kalshi, a regulated prediction market exchange. A contract pays one dollar if a specified team wins a game and zero otherwise. Its price can therefore be interpreted as a market-implied probability, subject to the usual caveats about heterogeneous beliefs, risk preferences, fees, bid--ask spreads, and market microstructure \citep{Manski2006,WolfersZitzewitz2006,GjerstadHall2005}. We merge one-minute Kalshi quotes, bid--ask spreads, volume, and open interest with timestamped NBA play-by-play data. The resulting dataset allows us to observe market prices, public signals, liquidity, and final payoffs at high frequency.

The setting provides a useful laboratory for three reasons. First, public information arrives continuously and is precisely recorded: every score, turnover, foul, timeout, rebound, possession, and clock movement changes the state of the game. Second, the payoff is simple, binary, and resolved within hours. Third, liquidity varies substantially across games and within games, allowing us to ask not only whether prices update, but when updating is more or less complete. This makes it possible to distinguish two notions that are often conflated. Prices may be directionally responsive, in the sense that they move in the right direction after public news, while still failing to update efficiently, in the sense that they do not move by the correct amount. Our empirical analysis proceeds in three steps. We first examine pre-game prices. Closing pre-game prices are well calibrated and become more accurate over the final 24 hours before tipoff. The pre-game price revision is strongly predictive of final outcomes, indicating that Kalshi prices contain meaningful information before the game begins. We then study live price responses to public signals. Prices react quickly and in the expected direction: favorable scoring, made shots, opponent turnovers, lead changes, and scoring runs all increase the market-implied probability of the referenced team winning. This establishes that the market processes public information in real time.
The stronger test compares price changes with changes in a benchmark win probability constructed from pre-game prices and live game states. Relative to this public-information benchmark, live prices underreact on impact. A one-minute change in benchmark win probability is associated with only about a 0.64-for-one contemporaneous change in the Kalshi midpoint. The missing adjustment predicts subsequent price drift over the next several minutes, including drift net of future changes in the benchmark probability. Thus, prices move in the right direction, but not far enough. The evidence is consistent with gradual incorporation of public information rather than instantaneous efficient updating. We then ask when this incomplete updating is most pronounced. The answer depends on the interaction between salience and liquidity. This interaction connects salience-based models of attention and choice \citep{BordaloGennaioliShleifer2012} with market-microstructure and limits-to-arbitrage mechanisms, in which trading costs and limited liquidity prevent immediate correction of price errors \citep{Kyle1985,GlostenMilgrom1985,AmihudMendelson1986,ShleiferVishny1997}. Salient public signals, such as three-point shots, lead changes, and scoring runs, are incorporated relatively quickly in liquid markets. However, when salient signals arrive in thin markets, prices adjust less completely on impact. The resulting underreaction gaps predict further price drift. Illiquidity therefore does not merely raise trading costs; it shapes the speed with which public information is incorporated into prices. At the same time, the predictable midpoint drift does not translate into a simple arbitrage opportunity once bid--ask costs are imposed. The evidence is best interpreted as gradual price discovery under trading frictions. 

The paper contributes to three literatures. First, it contributes to the literature on prediction markets. Existing work shows that prediction markets often aggregate information and produce accurate forecasts \citep{WolfersZitzewitz2004}, while also emphasizing that prices need not mechanically equal objective probabilities \citep{Manski2006,WolfersZitzewitz2006}. Calibration studies document that prediction markets can be informative but may display favorite--longshot patterns or other systematic deviations \citep{PageClemen2013,Page2012}. Recent work on Kalshi shows that prices are informative and become more accurate near resolution, while also displaying pricing patterns shaped by the platform's market design \citep{BurgiDengWhelan2026}. We shift the focus from static calibration to dynamic belief updating: not only whether prices are right on average, but whether they move correctly when public information arrives. Second, the paper contributes to the literature using sports betting and in-play prediction markets to test market efficiency \citep{ThalerZiemba1988,Sauer1998}. Closely related work studies high-frequency price responses around salient sports events. \citet{CroxsonReade2014} exploit goals scored on the cusp of half-time in soccer and find rapid and full adjustment. \citet{AngeliniDeAngelisSingleton2022} study in-play prediction markets around first goals and document both mispricing and behavioral patterns. \citet{GauriotPage2018} show that perceived momentum in football contests can affect behavior even when its underlying predictive content is limited, while \citet{GauriotPage2026} use knife-edge public information shocks in binary options markets and find mostly rapid adjustment, with short-lived underreaction to large shocks. Related evidence from second-by-second football betting markets shows that betting volume shifts toward teams that appear to gain momentum even when this perceived momentum is not associated with better outcomes \citep{OettingDeutscherSingletonDeAngelis2025}. Our setting differs because basketball generates a dense sequence of public signals within each game, and because we observe both exchange liquidity and repeated changes in the public-information state. This allows us to study not only whether prices respond to news, but how the completeness of updating varies with salience and trading conditions. Third, the paper contributes to behavioral finance research on underreaction, overreaction, attention, and limits to arbitrage. Traditional financial-market tests often face a joint-hypothesis problem because fundamental values are difficult to observe. Event contracts mitigate this problem: payoffs are binary, public signals are timestamped, and benchmark win probabilities can be estimated from comparable game states. The results show that even in a simple real-money market with transparent public information, prices may update gradually when attention and trading frictions interact.

The rest of the paper is organized as follows. Section~\ref{sec:data} describes Kalshi's institutional setting, the NBA contracts, and the construction of market-implied probabilities. Section~\ref{sec:framework} presents a framework for real-time belief updating. Section~\ref{sec:prematch} studies pre-game calibration and static biases. Section~\ref{sec:live} studies live belief updating around public information shocks. Section~\ref{sec:mechanisms} investigates mechanisms, focusing on salience, liquidity, and gradual adjustment. Section~\ref{sec:conclusion} concludes the paper.

\section{Institutional Setting and Data}
\label{sec:data}

We study National Basketball Association (NBA) winner contracts traded on Kalshi, a regulated event-contract exchange. A typical contract pays one dollar if the referenced team wins a specified NBA game and zero otherwise. Because payoffs are binary and resolved shortly after the game, contract prices can be interpreted as market-based forecasts of game outcomes, subject to the usual caveats regarding heterogeneous beliefs, risk preferences, trading costs, and market microstructure. Our market data consist of one-minute observations for NBA winner contracts. For each contract-minute, we observe the best YES bid, the best YES ask, trading volume, and open interest. Our main price measure is the midpoint between the best YES bid and best YES ask,
\[
p_{it}=\frac{bid_{it}+ask_{it}}{2},
\]
where (i) indexes contracts and (t) indexes time. We interpret ($p_{it}$) as the market-implied probability that the contract pays out. To characterize trading conditions, we measure liquidity using the bid--ask spread,
\[
spread_{it}=ask_{it}-bid_{it},
\]
together with minute-level trading volume and open interest. We merge the Kalshi data with NBA play-by-play records. The play-by-play feed provides a timestamped record of all publicly observable game events, including scores, turnovers, fouls, timeouts, rebounds, possessions, substitutions, and clock updates. Using these data, we reconstruct the public state of each game at every Kalshi observation. For each contract-minute, we assign the most recent observed game state and construct variables describing score margin, period, time remaining, recent scoring, made three-point shots, turnovers, lead changes, timeouts, and scoring runs. A key feature of the dataset is that all live variables are oriented toward the team referenced by the contract. Positive score margins, positive recent scoring runs, and positive event indicators always correspond to information favorable to the YES contract. This contract-oriented structure allows us to pool all contracts within a common empirical framework and interpret coefficients consistently across games and teams. The merge proceeds in two stages. We first match Kalshi contracts to NBA games using event dates, team identifiers, and NBA game identifiers. We then merge each contract's one-minute price series with the corresponding play-by-play file. The resulting dataset contains one observation per contract-minute and combines market prices, liquidity measures, public-information variables, and realized outcomes. Table~\ref{tab:sample_description} summarizes the final sample. The cleaned dataset contains 1,438 NBA games, 2,876 team-level contracts, and 409,512 contract-minute observations. The pre-game analysis uses 2,839 contracts from 1,421 games for which both the 24-hour pre-game price and the closing pre-game price are observed.

\begin{table}[h!]\centering
\caption{Sample description}\label{tab:sample_description}
\begin{tabular}{lc}
\toprule
Statistic & Value \\
\midrule
Games & 1,438 \\
Contracts & 2,876 \\
Contract-minute observations & 409,512 \\
Sample period & 2025-04-15 to 2026-05-25 \\
Pre-game 24h sample & 2,839 contracts from 1,421 games \\
Overtime games & 69 \\
Median game duration, real minutes & 138.6 \\
\bottomrule
\end{tabular}
\begin{minipage}{0.88\textwidth}
\vspace{0.1cm}\footnotesize Notes: The table summarizes the cleaned Kalshi--NBA dataset. A contract pays one dollar if the referenced team wins and zero otherwise. The Kalshi midpoint is the midpoint between the best YES bid and best YES ask. Live observations are measured at the contract-minute level. The 24-hour pre-game sample includes contracts for which both the price 24 hours before game start and the last pre-game price are observed.
\end{minipage}
\end{table}

The dataset offers several advantages for studying market efficiency and belief updating. First, the traded claim has a transparent binary payoff. Second, public information arrives continuously and is observed at high frequency through the play-by-play feed. Third, prices and liquidity measures are observed at the same frequency as the information flow. Finally, outcomes are realized within hours rather than months or years. This combination allows us to study not only whether prediction market prices are informative, but also how rapidly and completely they incorporate public information in real time.

\section{A Framework for Real-Time Belief Updating}
\label{sec:framework}

This section develops a simple framework for interpreting the empirical analysis. The objective is not to estimate a structural model of trading, but to discipline the distinction between three objects: the statistical content of public information, the market price response, and subsequent price correction. Consider a binary contract \(i\) that pays one dollar if the referenced team wins the game and zero otherwise. Let \(Y_i\in\{0,1\}\) denote the terminal payoff. At time \(t\), market participants observe a public information set \(\mathcal I_{it}\), which includes pre-game information and the live game state. The public-information benchmark is
\begin{equation} \label{eq:q}
    q_{it}=\Pr(Y_i=1\mid \mathcal I_{it}).
\end{equation}

This is the probability of winning implied by the observed state of the game. It is not observed by the econometrician, but can be approximated using a win-probability model estimated from historical game states. Let \(p_{it}\) denote the Kalshi midpoint, interpreted as the market-implied probability that the contract pays out. The efficient-updating benchmark is
\begin{equation} \label{eq:p}
   p_{it}=q_{it}+\varepsilon_{it}
\end{equation}
where \(\varepsilon_{it}\) is an unpredictable pricing error. In levels, this implies calibration. In changes, it implies one-for-one updating:
\begin{equation} \label{eq:deltap}
\Delta p_{it}=\Delta q_{it}+\eta_{it},
\end{equation}
where
\[
\Delta p_{it}=p_{it}-p_{i,t-1},
\qquad
\Delta q_{it}=q_{it}-q_{i,t-1}.
\]
Under efficient updating, public information should be incorporated immediately, and the residual component \(\eta_{it}\) should not predict future price changes. We allow for the possibility that prices incorporate public information gradually or with behavioral distortions. Let
\begin{equation} \label{eq:m}
m_{it}=p_{it}-q_{it}
\end{equation}
denote mispricing relative to the public-information benchmark. A parsimonious updating equation is
\begin{equation} \label{eq:deltap1}
\Delta p_{it}
=
\lambda_{it}\Delta q_{it}
+
\rho m_{i,t-1}
+
\eta_{it}.
\end{equation}
The parameter \(\lambda_{it}\) captures the contemporaneous intensity of updating. If \(\lambda_{it}=1\), prices move one-for-one with the benchmark probability. If \(\lambda_{it}<1\), prices underreact on impact. If \(\lambda_{it}>1\), prices overreact. The parameter $(\rho)$ captures correction or persistence of previous mispricing. If prices gradually correct past underreaction, lagged mispricing should predict subsequent price movements. We allow updating intensity to depend on signal and market characteristics:
\begin{equation} \label{eq:lambda}
\lambda_{it}
=
1
+
\alpha_S Salience_{it}
+
\alpha_L Illiquidity_{it}
+
\alpha_{SL} Salience_{it}\cdot Illiquidity_{it}.
\end{equation}
Salience captures whether the public signal is especially visible or attention-grabbing, such as a made three-point shot, a lead change, a large scoring run, or a late-game event. Illiquidity captures the cost of trading and is measured using bid--ask spreads, recent trading volume, and open interest. Substituting into the updating equation gives
\begin{equation} \label{eq:deltap2}
\Delta p_{it}
=
\Delta q_{it}
+
\alpha_S Salience_{it}\Delta q_{it}
+
\alpha_L Illiquidity_{it}\Delta q_{it}
+
\alpha_{SL}
\left(Salience_{it}\cdot Illiquidity_{it}\right)\Delta q_{it}
+
\rho m_{i,t-1}
+
\eta_{it}.
\end{equation}
This equation nests the efficient benchmark. Efficient real-time updating requires one-for-one incorporation of \(\Delta q_{it}\), no systematic dependence of the response on salience or liquidity, and no predictable subsequent drift. Deviations from these restrictions generate the empirical Hypothesis below.

\paragraph{Hypothesis 1: Pre-game informativeness.}
If prices aggregate information before the game, pre-game prices should predict realized outcomes. In particular,
$\mathbb E[Y_i\mid p_{i,0}]$
should be increasing in \(p_{i,0}\), and calibration implies an intercept close to zero and a slope close to one in regressions of \(Y_i\) on \(p_{i,0}\).

\paragraph{Hypothesis 2: Directional live updating.}
If live prices respond to public information, favorable benchmark innovations should increase prices and unfavorable innovations should decrease them:
\[
\mathbb E[\Delta p_{it}\mid \Delta q_{it}>0]>0,
\qquad
\mathbb E[\Delta p_{it}\mid \Delta q_{it}<0]<0.
\]
This is a weak form of information aggregation: prices move in the right direction, but not necessarily by the right amount.

\paragraph{Hypothesis 3: Efficient live updating.}
Efficient updating requires \eqref{eq:deltap}.
We test this prediction using
\begin{equation}
\label{eq:deltap_emp}
\Delta p_{it}
=
\alpha
+
\beta \Delta q_{it}
+
\Gamma X_{it}
+
\eta_{it},
\end{equation}
where \(X_{it}\) contains state and market controls. Under efficient updating, market prices should move one-for-one with changes in the benchmark probability, implying
$\beta=1$.
Values of \(\beta<1\) indicate underreaction, while values of \(\beta>1\) indicate overreaction.

\paragraph{Hypothesis 4: Gradual correction of updating errors.}
If prices do not fully incorporate public information on impact, either because of underreaction or overreaction, the resulting pricing error should predict subsequent corrections. Define the updating gap as
\begin{equation}
\label{eq:gap}
Gap_{it}=\Delta q_{it}-\Delta p_{it}.
\end{equation}
A positive gap indicates underreaction, while a negative gap indicates overreaction. If prices gradually converge toward the benchmark probability, then future price changes should be related to the size and sign of the gap:
\begin{equation}
\label{eq:drift}
p_{i,t+h}-p_{it}
=
\alpha_i+\delta_t+\rho Gap_{it}+\varepsilon_{i,t+h}.
\end{equation}

Under gradual correction, the prediction is
$\rho>0$.
A positive coefficient implies that positive gaps (underreaction) are followed by upward price adjustments, whereas negative gaps (overreaction) are followed by downward price adjustments.

A stronger test nets out subsequent changes in the benchmark probability:
\begin{equation}
\label{eq:drift1}
(p_{i,t+h}-p_{it})-(q_{i,t+h}-q_{it})
=
\alpha_i+\delta_t+\rho Gap_{it}+\varepsilon_{i,t+h}.
\end{equation}
A positive value of \(\rho\) in this specification indicates that pricing errors are corrected over time, beyond the arrival of new public information.

\paragraph{Hypothesis 5: Salience affects the completeness of updating.}
If salient signals attract attention or distort beliefs, the price response should differ systematically following salient public information. In the updating equation \eqref{eq:deltap2}, this corresponds to
$\alpha_S \neq 0$.

A non-zero value of \(\alpha_S\) implies that salience affects the sensitivity of prices to benchmark probability changes. Positive values indicate stronger updating, whereas negative values indicate weaker updating relative to non-salient events.

\paragraph{Hypothesis 6: Liquidity disciplines real-time updating.}
If trading frictions affect price adjustment, deviations from efficient updating should be more pronounced when markets are illiquid. The key hypothesis is that the effect of salience depends on liquidity:
$\alpha_{SL}\neq 0$ in \eqref{eq:deltap2}.
A non-zero interaction effect implies that liquidity moderates the impact of salient information on price updating. In particular, salient events may generate larger deviations from efficient updating when market liquidity is low.

\bigskip

These Hypotheses organize the empirical analysis. Section~\ref{sec:prematch} tests whether pre-game prices provide an informative prior. Section~\ref{sec:live} tests whether live prices move in the right direction and whether they update one-for-one with benchmark win-probability changes. Section~\ref{sec:mechanisms} studies whether incomplete updating is concentrated in salient and illiquid states and whether updating errors predict subsequent price corrections.

\section{Pre-Game Calibration}
\label{sec:prematch}

\begin{figure}[h!]
\centering
\includegraphics[width=0.78\textwidth]{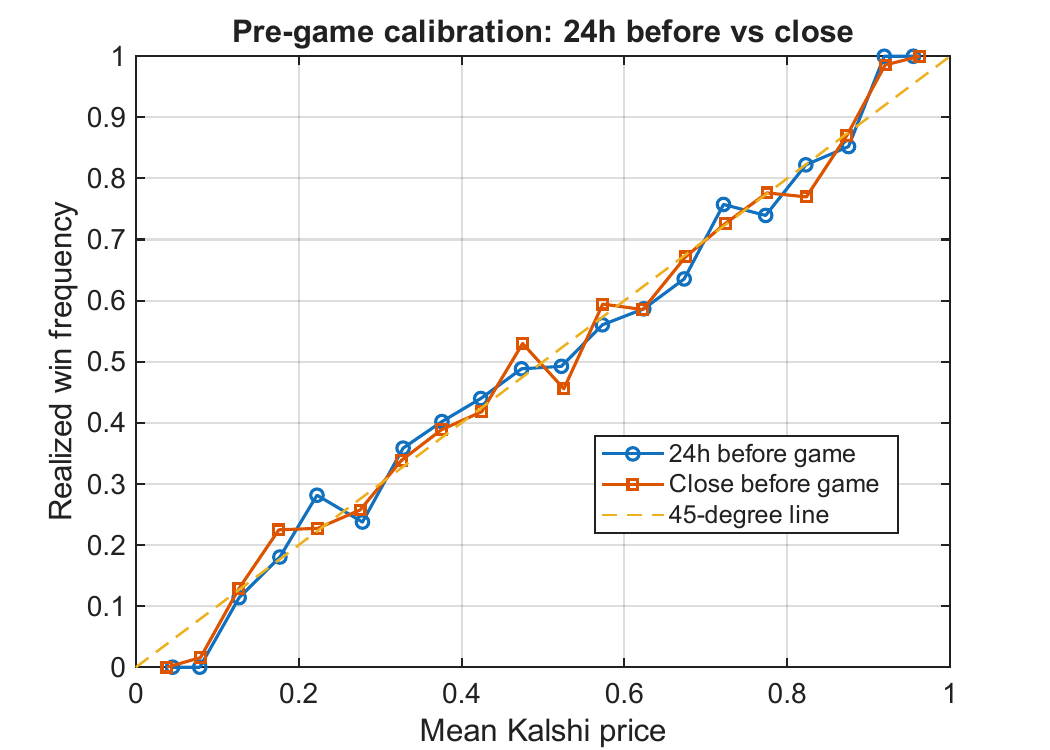}
\caption{Pre-game calibration: 24 hours before game start and closing prices}
\label{fig:prematch_calibration_24h_close}
\begin{minipage}{0.88\textwidth}
\vspace{0.1cm}
\footnotesize
Notes: The figure compares Kalshi-implied probabilities with realized win frequencies. Contracts are grouped into five-percentage-point bins. The 24-hour price is the last observed midpoint at least 24 hours before game start. The closing price is the last observed midpoint before the first NBA play-by-play event.
\end{minipage}
\end{figure}

We begin by studying pre-game prices. This exercise provides a baseline for the live analysis. If pre-game prices were uninformative, subsequent tests of real-time updating would be difficult to interpret. Conversely, if pre-game prices are calibrated and become more accurate before tipoff, live price movements can be interpreted as updates from a meaningful market prior. 
We define two pre-game prices. The first, \(p_{i,-24h}\), is the last observed Kalshi midpoint at least 24 hours before game start. The second, \(p_{i,0}\), is the last observed midpoint before the first NBA play-by-play event. The 24-hour revision is
\[
\Delta p_i^{24h}=p_{i,0}-p_{i,-24h}.
\]
Our 24-hour pre-game sample contains 2,839 contracts from 1,421 games. Figure~\ref{fig:prematch_calibration_24h_close} reports calibration plots for prices observed 24 hours before tipoff and for closing pre-game prices. Contracts are grouped into five-percentage-point price bins, and the figure compares the average market-implied probability in each bin with the realized win frequency. Prices are informative at both horizons, and closing prices lie close to the 45-degree line.

Table~\ref{tab:prematch_24h_update} shows that forecast accuracy improves over the final 24 hours before the game. The mean Brier score falls from 0.204 at the 24-hour horizon to 0.199 at the close. The improvement of 0.0046 is statistically significant, with standard errors clustered by game. The mean absolute error also declines, from 0.406 to 0.397. These magnitudes are economically modest but statistically precise, consistent with gradual information aggregation as tipoff approaches.

\begin{table}[h!]
\centering
\caption{Pre-game updating over the final 24 hours}
\label{tab:prematch_24h_update}
\begin{tabular}{lccc}
\toprule
 & 24h before & Close & Improvement \\
\midrule
\multicolumn{4}{l}{\textit{Panel A. Forecast accuracy}} \\
Mean Brier score & 0.204 & 0.199 & 0.0046*** \\
 &  &  & (0.0013) \\
Mean absolute error & 0.406 & 0.397 & 0.0086*** \\
 &  &  & (0.0014) \\
\addlinespace
\multicolumn{4}{l}{\textit{Panel B. Calibration and information in revisions}} \\
Price 24h before game & 0.977 &  & 0.964 \\
 & (0.046) &  & (0.046) \\
Closing pre-game price &  & 0.985*** &  \\
 &  & (0.044) &  \\
Revision, close minus 24h &  &  & 1.292*** \\
 &  &  & (0.217) \\
Constant & 0.009 & 0.006 & 0.017 \\
 & (0.023) & (0.022) & (0.023) \\
\midrule
Observations & 2,839 & 2,839 & 2,839 \\
Games & 1,421 & 1,421 & 1,421 \\
\bottomrule
\end{tabular}
\begin{minipage}{0.88\textwidth}
\vspace{0.1cm}
\footnotesize
Notes: The table compares the last observed midpoint at least 24 hours before game start with the last pre-game midpoint. In Panel A, improvement is defined as the 24-hour forecast error minus the closing forecast error, so positive values indicate that the closing price is more accurate. In Panel B, the dependent variable is an indicator equal to one if the contract pays out. Standard errors, in parentheses, are clustered by game. 
For the coefficients on pre-game prices, statistical significance is based on tests of the null hypothesis $H_0:\beta=1$.
*, **, and *** denote significance at 10\%, 5\%, and 1\% levels, respectively.
\end{minipage}
\end{table}

The pre-game revision is itself informative. Column 3 of Table \ref{tab:prematch_24h_update} estimates
\begin{equation} \label{eq:pre}
  Y_i=\alpha+\beta p_{i,-24h}+\gamma
\Delta p_i^{24h}+\varepsilon_i.  
\end{equation}
The coefficient on the revision is 1.29 and highly statistically significant. Conditional on the 24-hour price, upward revisions predict higher realized win probabilities and downward revisions predict lower realized win probabilities. A 10 percentage point increase in the pre-game revision is associated with a roughly 13 percentage point increase in the probability that the contract pays out. The evidence does not point to large static mispricing. The calibration regressions show intercepts close to zero and slopes close to one at both horizons. Bin-level excess payoffs display some variation across the price distribution, especially in sparsely populated extreme bins, but the dominant pattern is that prices are informative and become more accurate as the market approaches game start. This result establishes the starting point for the live analysis. Kalshi prices provide a meaningful pre-game prior. The central question is therefore not whether the market has information before the game starts, but whether it updates this prior efficiently when public information begins to arrive in real time.

\section{Live Belief Updating}
\label{sec:live}

\begin{figure}[h!]
\centering
\includegraphics[width=0.78\textwidth]{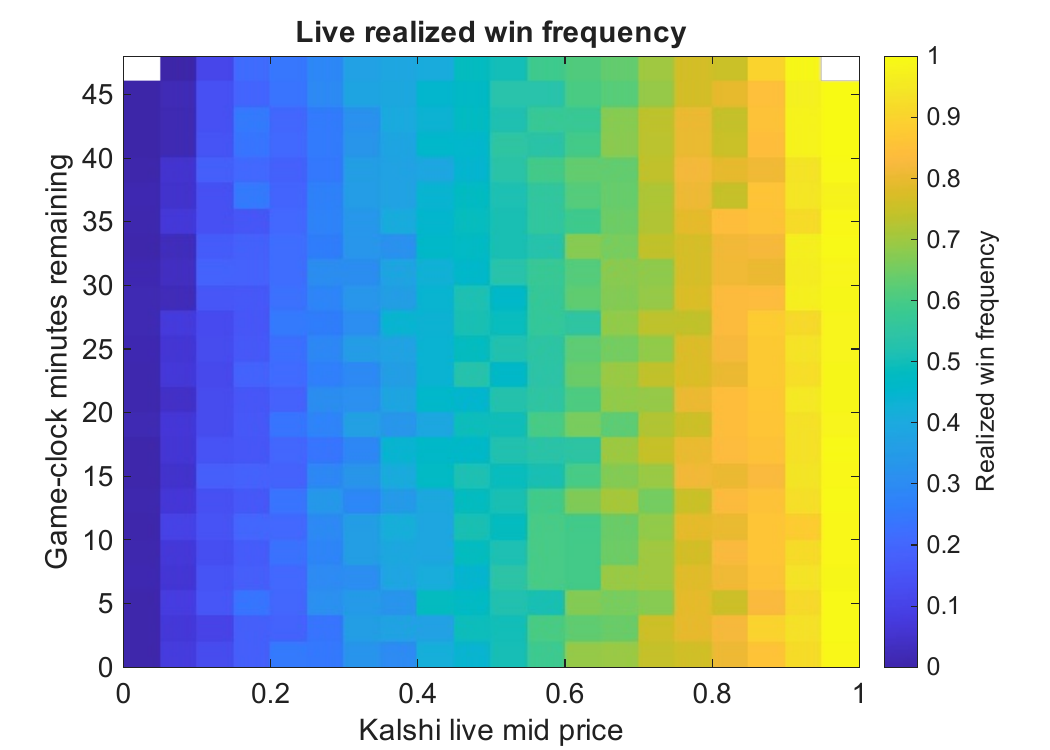}
\caption{Live calibration by price and game-clock time remaining}
\label{fig:live_calibration_gameclock}
\begin{minipage}{0.88\textwidth}
\vspace{0.1cm}
\footnotesize
Notes: The figure reports realized win frequencies by bins of the Kalshi live midpoint and game-clock minutes remaining. The unit of observation is a contract-minute. Cells with limited support are left blank. The figure is descriptive and does not condition on the full game state.
\end{minipage}
\end{figure}

Having established that pre-game prices are informative, we next study how prices respond once public information begins to arrive during the game. This section tests the weakest implication of real-time information aggregation: favorable public signals should increase the price of the contract, while unfavorable signals should decrease it. This test does not require a benchmark win-probability model and does not ask whether the response is of the correct magnitude. It asks only whether live prices move in the right direction. We first provide descriptive evidence on live calibration. Figure~\ref{fig:live_calibration_gameclock} plots realized win frequencies across bins of the Kalshi live midpoint and game-clock minutes remaining. The figure pools contract-minutes across games and does not condition on the full game state, so it should be interpreted descriptively. Nevertheless, live prices are strongly informative throughout the game: higher prices correspond to higher realized win frequencies, and the relationship becomes sharper as the game approaches resolution.

To study directional updating, we construct signed public-signal variables. All signals are oriented toward the team referenced by the contract. For example, \(signed\_points\_1m\) equals points scored by the contract team over the previous minute minus points scored by the opponent. Similarly, \(signed\_3pt\) equals one if the contract team made a three-point shot in the previous minute and minus one if the opponent did. For turnovers, the sign is reversed: an opponent turnover is good news for the contract team, while a turnover by the contract team is bad news. Thus, positive values always correspond to information favorable to the YES contract. The outcome is the one-minute change in the Kalshi midpoint,
$\Delta p_{it}$.
Table~\ref{tab:live_directional_means} reports mean price changes by signal direction. The pattern is highly symmetric. When the contract team outscores the opponent in the previous minute, the price rises by 2.8 percentage points on average; when the opponent outscores the contract team, the price falls by 2.8 percentage points. A made three-point shot by the contract team is associated with an increase of 3.7 percentage points, while an opponent three-point shot is associated with a decline of 3.8 percentage points. Lead changes generate even larger directional movements, around 4.5 percentage points in the expected direction. Sustained scoring runs also convey economically meaningful information. An 8--0 run is associated with an average price change of about 2.5 percentage points, increasing to 2.8 percentage points for a 10--0 run, with nearly identical effects of opposite sign when the run is achieved by the opposing team.

\begin{table}[h!]\centering
\caption{Live price changes by public-signal direction}
\label{tab:live_directional_means}
\begin{tabular}{lccc}
\toprule
Signal & Negative signal & No signal & Positive signal \\
\midrule
Net points, last 1 min & -0.0284 & 0.0001 & 0.0281 \\
Net points, last 5 min & -0.0125 & 0.0000 & 0.0124 \\
Made 3pt & -0.0377 & 0.0000 & 0.0370 \\
Made 2pt & -0.0244 & 0.0000 & 0.0240 \\
Turnover & -0.0226 & 0.0000 & 0.0223 \\
Lead change & -0.0455 & 0.0000 & 0.0451 \\
Run 8--0 & -0.0250 & 0.0000 & 0.0249 \\
Run 10--0 & -0.0280 & 0.0000 & 0.0279 \\
\bottomrule
\end{tabular}
\begin{minipage}{0.88\textwidth}
\vspace{0.1cm}
\footnotesize
Notes: The table reports mean one-minute changes in the Kalshi midpoint by signal direction. Signals are signed so that positive values correspond to information favorable to the team referenced by the contract. The unit of observation is a contract-minute.
\end{minipage}
\end{table}

We next estimate regressions of the form
\begin{equation} \label{eq:signal}
\Delta p_{it}
=
\alpha
+
\beta Signal_{it}
+
\Gamma X_{it}
+
\varepsilon_{it},
\end{equation}
where \(X_{it}\) includes the lagged midpoint, score margin, absolute score margin, game-clock minutes remaining, bid--ask spread, recent volume, and open interest. Standard errors are clustered by game. Table~\ref{tab:live_directional_regressions} confirms the descriptive evidence. Recent scoring by the contract team predicts immediate price increases. One net point scored in the previous minute raises the Kalshi midpoint by 1.22 percentage points. Event-level signals also move prices in the expected direction: made three-point shots, made two-point shots, opponent turnovers, and lead changes all generate positive and statistically significant responses. Salient events, including lead changes, made three-point shots, 8--0 scoring runs, and the signed salience index, also predict immediate price movements in the expected direction.

\begin{table}[h!]\centering
\caption{Directional live updating}
\label{tab:live_directional_regressions}
\begin{tabular}{lccc}
\toprule
 & (1) & (2) & (3) \\
 & Scoring & Event types & Salience \\
\midrule
Net points, last 1 min & 0.0122*** &  &  \\
 & (0.0001) &  &  \\
Net points, last 5 min & 0.0010*** &  &  \\
 & (0.0000) &  &  \\
Made 3pt &  & 0.0372*** & 0.0264*** \\
 &  & (0.0005) & (0.0004) \\
Made 2pt &  & 0.0244*** &  \\
 &  & (0.0003) &  \\
Turnover &  & 0.0102*** &  \\
 &  & (0.0002) &  \\
Lead change &  & 0.0178*** & 0.0240*** \\
 &  & (0.0007) & (0.0008) \\
Run 8--0 &  &  & 0.0048*** \\
 &  &  & (0.0009) \\
Salience &  &  & 0.0071*** \\
 &  &  & (0.0002) \\
\midrule
Controls & Yes & Yes & Yes \\
Observations & 362,133 & 362,133 & 362,133 \\
Game clusters & 1,438 & 1,438 & 1,438 \\
\bottomrule
\end{tabular}
\begin{minipage}{0.90\textwidth}
\vspace{0.1cm}
\footnotesize
Notes: The dependent variable is the one-minute change in the Kalshi live midpoint. Signals are signed so that positive values correspond to information favorable to the team referenced by the contract. Controls include the lagged midpoint, score margin, absolute score margin, game-clock minutes remaining, bid--ask spread, recent volume, and open interest. Standard errors, in parentheses, are clustered by game. *, **, and *** denote significance at 10\%, 5\%, and 1\% levels, respectively.
\end{minipage}
\end{table}

The estimates show that live prices are not detached from the game state. Traders process public information quickly, and prices move sharply in the expected direction after favorable and unfavorable signals. This supports directional live updating. However, directional responsiveness is not the same as efficient updating. A market can move in the right direction and still move too little or too much relative to the signal's statistical content. To test whether prices move by the right amount, the next section compares Kalshi price changes with changes in a benchmark win probability implied by the public game state.

\subsection{Benchmark Win Probabilities and Efficient Updating}
\label{subsec:benchmark_updating}

The evidence above shows that live prices move in the right direction after public signals. Directional updating, however, is only a weak requirement for efficiency. A market can respond positively to favorable news and still move too little or too much relative to the news' statistical content. We therefore compare Kalshi price changes with changes in a benchmark win probability implied by the public game state. 

\begin{figure}[h!]
\centering
\includegraphics[width=0.78\textwidth]{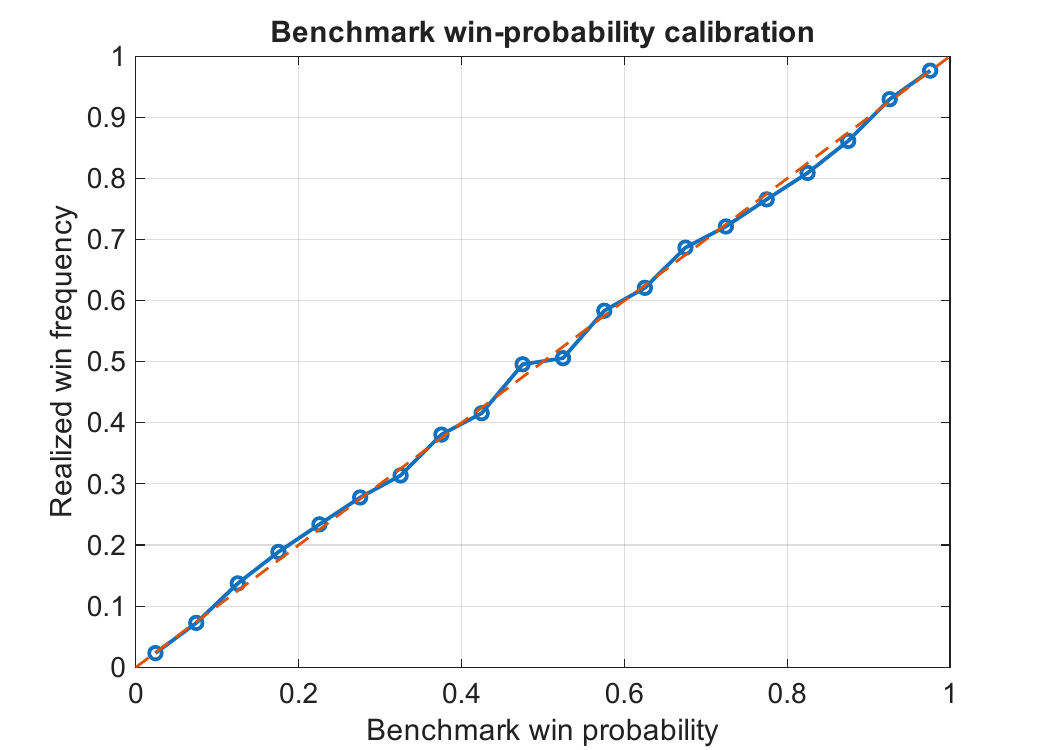}
\caption{Calibration of the benchmark win-probability model}
\label{fig:qhat_calibration}
\begin{minipage}{0.88\textwidth}
\vspace{0.1cm}
\footnotesize
Notes: The figure compares out-of-sample benchmark win probabilities with realized win frequencies. The benchmark is estimated using five-fold cross-fitting at the game level and excludes contemporaneous Kalshi live prices. Contracts are grouped into five-percentage-point bins.
\end{minipage}
\end{figure}

Following Section~\ref{sec:framework}, we estimate the benchmark probability
$q_{it}$
using a logit model of the final payoff on pre-game and live game-state variables, including the pre-game closing price, score margin, absolute score margin, game-clock time remaining, period indicators, home status, recent net scoring, and nonlinear interactions between score margin and time remaining. The benchmark excludes the contemporaneous Kalshi live price. This ensures that we compare market prices with an independent public-information benchmark rather than with a model that mechanically embeds the market price. To reduce overfitting, we estimate the benchmark out of sample using five-fold cross-fitting at the game level. All observations from a given game are assigned to the same fold. For each fold, the model is estimated on the remaining games and used to predict \(q_{it}\) for the held-out games. This procedure produces an out-of-sample benchmark probability for every contract-minute. The benchmark is well calibrated. Its Brier score is 0.164, compared with 0.164 for the Kalshi live midpoint and 0.211 for the pre-game closing price. Figure~\ref{fig:qhat_calibration} shows that realized win frequencies closely track benchmark probabilities across the distribution. The benchmark therefore provides a useful proxy for the public-information component of win probability.

We next test the efficient-updating condition in Equation~\eqref{eq:deltap} using the empirical specification in Equation~\eqref{eq:deltap_emp}. Table~\ref{tab:efficient_live_updating} reports the results. Column 1 estimates the relationship between Kalshi price changes and benchmark probability changes without additional controls. Column 2 adds the control variables \(X_{it}\), which include the lagged Kalshi midpoint, the lagged benchmark probability, score margin, absolute score margin, game-clock minutes remaining, bid--ask spread, recent volume, and open interest.

Column 1 shows the raw relationship between Kalshi price changes and benchmark probability changes. The coefficient on \(\Delta q_{it}\) is 0.630. Column 2 adds controls and yields a similar estimate of 0.638. In both cases, the coefficient is precisely estimated and statistically different from one. Thus, a 10 percentage point increase in benchmark win probability is associated with only about a 6.4 percentage point contemporaneous increase in the Kalshi midpoint. The evidence therefore points to incomplete incorporation of public information on impact.

\begin{table}[h!]\centering
\caption{Benchmark probability changes and live price updating}
\label{tab:efficient_live_updating}
\begin{tabular}{lcc}
\toprule
 & (1) & (2) \\
 & \(\Delta p_{it}\) & \(\Delta p_{it}\) \\
\midrule
\(\Delta q_{it}\) & 0.630*** & 0.638*** \\
 & (0.005) & (0.010) \\
Lagged midpoint &  & -0.079** \\
 &  & (0.038) \\
Lagged benchmark probability &  & 0.078* \\
 &  & (0.040) \\
Score margin &  & 0.0001 \\
 &  & (0.0001) \\
Minutes remaining &  & 0.0000*** \\
 &  & (0.0000) \\
Spread &  & 0.0003 \\
 &  & (0.0009) \\
Recent volume &  & Yes \\
Open interest &  & Yes \\
\midrule
Observations & 356,769 & 356,769 \\
Game clusters & 1,438 & 1,438 \\
\(p\)-value (\(H_0:\beta=1\)) & \(<0.001\) & \(<0.001\) \\
\bottomrule
\end{tabular}
\begin{minipage}{0.90\textwidth}
\vspace{0.1cm}
\footnotesize
Notes: The dependent variable is the one-minute change in the Kalshi live midpoint. The benchmark probability is estimated out of sample using five-fold cross-fitting at the game level. Controls include the lagged midpoint, lagged benchmark probability, score margin, absolute score margin, game-clock minutes remaining, bid--ask spread, recent volume, and open interest. Standard errors are clustered by game. ***, **, and * denote significance at the 1, 5, and 10 percent levels, respectively.
\end{minipage}
\end{table}

A natural concern is that measurement error in the benchmark probability could attenuate the estimated coefficient on benchmark probability changes. We therefore examine whether the updating gap defined in Equation~\eqref{eq:gap} predicts subsequent price adjustments. Following Hypothesis~4, we estimate Equations~\eqref{eq:drift} and \eqref{eq:drift1} for horizons \(h=1,2,5,10,\) and \(15\) minutes. As in the updating regressions, the control variables include the lagged midpoint, lagged benchmark probability, score margin, absolute score margin, game-clock minutes remaining, bid--ask spread, recent volume, and open interest. Table~\ref{tab:underreaction_drift} reports the results.
The coefficient on the updating gap is positive and highly statistically significant at every horizon. At the five-minute horizon, a 10 percentage point initial updating gap predicts a 2.0 percentage point subsequent price change and a 4.6 percentage point price change net of benchmark updates. At the fifteen-minute horizon, the corresponding effects are 2.4 and 4.8 percentage points. These results indicate that deviations from efficient updating are subsequently corrected rather than permanently incorporated into prices.

\begin{table}[h!]\centering
\caption{Future price drift after incomplete updating}
\label{tab:underreaction_drift}
\begin{tabular}{lcc}
\toprule
Horizon & Raw future price drift & Net of future benchmark changes \\
\midrule
1 min  & 0.150*** & 0.379*** \\
       & (0.011) & (0.019) \\
2 min  & 0.164*** & 0.414*** \\
       & (0.013) & (0.025) \\
5 min  & 0.195*** & 0.459*** \\
       & (0.016) & (0.029) \\
10 min & 0.196*** & 0.458*** \\
       & (0.018) & (0.031) \\
15 min & 0.236*** & 0.484*** \\
       & (0.019) & (0.032) \\
\midrule
Observations & 312,460 & 353,080 \\
Game clusters & 1,437 & 1,438 \\
\bottomrule
\end{tabular}
\begin{minipage}{0.90\textwidth}
\vspace{0.1cm}
\footnotesize
Notes: The table reports estimates of \(\rho\) from Equations~\eqref{eq:drift} and \eqref{eq:drift1}. The first column reports raw future price drift, while the second nets out subsequent changes in the benchmark probability. All regressions include the lagged midpoint, lagged benchmark probability, score margin, absolute score margin, game-clock minutes remaining, bid--ask spread, recent volume, and open interest. Standard errors are clustered by game. *, **, and *** denote significance at 10, 5, and 1 percent levels, respectively.
\end{minipage}
\end{table}

\begin{figure}[h!]
\centering
\includegraphics[width=0.78\textwidth]{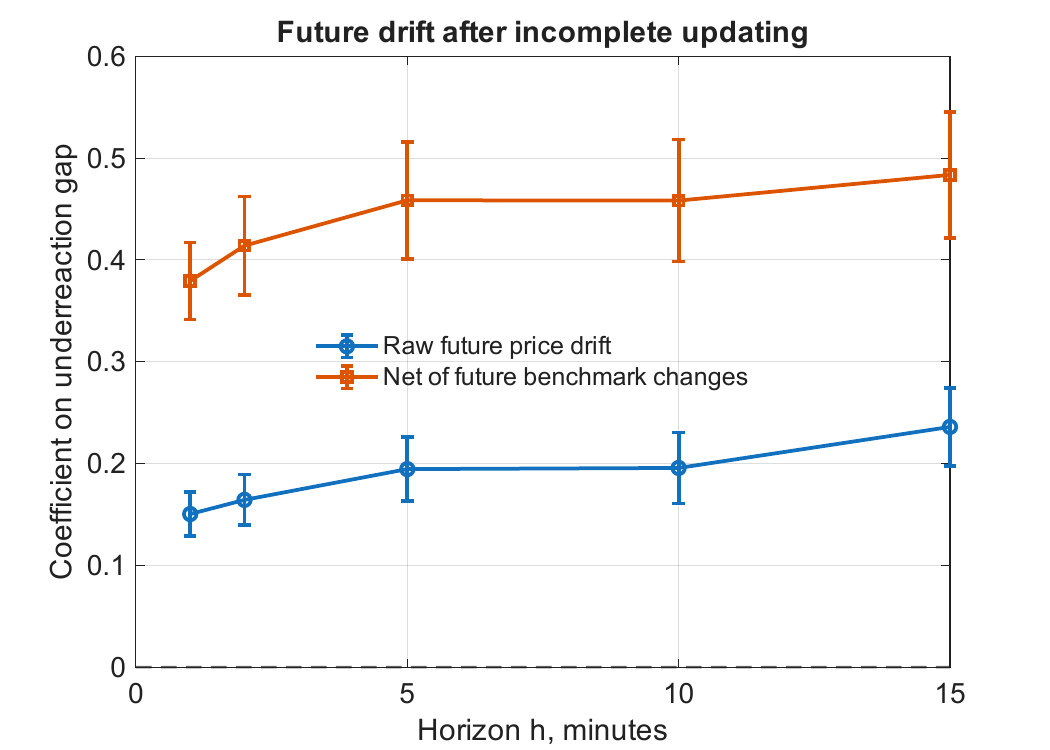}
\caption{Future price drift after incomplete updating}
\label{fig:underreaction_drift}
\begin{minipage}{0.88\textwidth}
\vspace{0.1cm}
\footnotesize
Notes: The figure reports estimates of \(\rho\) from Equations~\eqref{eq:drift} and \eqref{eq:drift1}. Vertical bars denote 95 percent confidence intervals. Standard errors are clustered by game.
\end{minipage}
\end{figure}

Figure~\ref{fig:underreaction_drift} summarizes the drift coefficients. The effect appears immediately and persists over the following fifteen minutes. The net-of-benchmark specification is especially informative: when the benchmark indicates that the contemporaneous price response was smaller than the benchmark-implied probability change, subsequent prices continue to move in the benchmark direction even after accounting for future public-information changes. This strengthens the interpretation that the one-minute response in Table~\ref{tab:efficient_live_updating} reflects gradual incorporation of public information rather than only noise in the estimated benchmark.

\ref{app:benchmark_models} shows that the result is robust to alternative benchmark win-probability models, including parsimonious, flexible, no-recent-scoring, and chronological-holdout specifications. \ref{app:microstructure} shows that the result is also robust to microstructure-related sample restrictions. The coefficient on benchmark probability changes remains well below one in samples with positive trading volume, narrow spreads, non-stale quotes, and high-quality quote observations. The updating gap also continues to predict five-minute drift net of future benchmark changes. However, executable-style returns that buy at the ask and sell at the bid are negative, indicating that the predictable midpoint drift is largely absorbed by trading costs. Taken together, the evidence supports a gradual-adjustment interpretation. Live prices respond strongly to public-information shocks, but the response is incomplete on impact. The initial updating gap predicts subsequent drift, including drift relative to future changes in the benchmark probability. The next section asks whether this incomplete updating is uniform, or whether it is concentrated after salient public signals and in thin markets. \ref{app:game_phase} shows that incomplete updating is present throughout the game and is especially pronounced in clutch situations, where the coefficient on benchmark probability changes falls to approximately 0.51.

\section{Salience, Liquidity, and Gradual Adjustment}
\label{sec:mechanisms}

The previous section shows that live prices respond strongly to public information but fail to incorporate it fully on impact. Price changes move in the same direction as benchmark win-probability changes, yet the contemporaneous response is substantially below one-for-one, and the resulting updating gap predicts future price drift. We now ask whether this incomplete updating is uniform across public signals and market states. The framework in Section~\ref{sec:framework} highlights two forces that may shape the speed of price discovery. First, some public signals are more salient than others. A made three-point shot, a lead change, or a large scoring run is likely to attract attention and may therefore be processed more rapidly than less visible information. Second, even when information is recognized, incorporating it into prices may be costly when liquidity is limited. Attention and liquidity therefore need not have the same effect: salient signals may be easier to notice, while liquidity determines how fully they are incorporated into prices. 
To study these mechanisms, we use the updating gap defined in Equation~\eqref{eq:gap}. A positive gap indicates that the benchmark probability increased more than the Kalshi price. To treat positive and negative benchmark innovations symmetrically, we define directional underreaction as
\[
UR_{it}
=
\operatorname{sign}(\Delta q_{it})
\left(\Delta q_{it}-\Delta p_{it}\right).
\]
Thus, ($UR_{it}>0$) indicates that the price moved too little in the benchmark-implied direction, while ($UR_{it}<0$) indicates that the price moved too much. We restrict attention to economically meaningful benchmark innovations, ($|\Delta q_{it}|\geq 0.0025$), to avoid classifying negligible probability changes as directional shocks. We measure salience using a standardized index constructed from highly visible game events, including made three-point shots, lead changes, scoring runs, and the event-level salience score derived from the play-by-play data. Illiquidity is measured using a standardized index that increases with bid--ask spreads and decreases with recent trading volume and open interest. Both indices are standardized to have mean zero and unit variance. We estimate a reduced-form version of Equation~\eqref{eq:deltap2} in which the dependent variable is directional underreaction,
\[
UR_{it}
=
\alpha
+
\beta Salience_{it}
+
\gamma Illiquidity_{it}
+
\theta Salience_{it}\cdot Illiquidity_{it}
+
\Gamma X_{it}
+
\varepsilon_{it},
\]
where $X_{it}$ includes the absolute benchmark innovation, lagged market and benchmark probabilities, score margin, absolute score margin, game-clock minutes remaining, and close-game and clutch indicators. Standard errors are clustered by game. Table~\ref{tab:mechanisms_impact} reports the results. Column 1 uses the composite salience index. Illiquidity is associated with greater directional underreaction: when markets are thinner, prices move less completely in the direction implied by the benchmark. By contrast, the coefficient on salience is negative. Salient events are associated with smaller underreaction gaps, suggesting that highly visible information is incorporated more rapidly when trading conditions are favorable. The interaction between salience and illiquidity is positive and highly significant. The effect of salience therefore depends on market conditions. In liquid markets, salient information is incorporated relatively quickly. In illiquid markets, however, the same salient signals are associated with larger underreaction gaps. Salience appears to improve recognition of public information, while liquidity determines whether that information is fully incorporated into prices.

\begin{table}[h!]\centering
\caption{Salience, liquidity, and impact underreaction}
\label{tab:mechanisms_impact}
\begin{tabular}{lcc}
\toprule
 & (1) & (2) \\
 & Composite salience & Event-specific salience \\
\midrule
Salience & -0.0026*** &  \\
 & (0.0001) &  \\
Illiquidity & 0.0015*** & 0.0008*** \\
 & (0.0002) & (0.0002) \\
Salience $\cdot$ Illiquidity & 0.0014*** &  \\
 & (0.0002) &  \\
Any 3pt &  & -0.0055*** \\
 &  & (0.0003) \\
Any turnover &  & -0.0064*** \\
 &  & (0.0002) \\
Any lead change &  & -0.0060*** \\
 &  & (0.0005) \\
Any run 8--0 &  & -0.0102*** \\
 &  & (0.0008) \\
Any 3pt $\cdot$ Illiquidity &  & 0.0027*** \\
 &  & (0.0003) \\
Lead change $\cdot$ Illiquidity &  & 0.0043*** \\
 &  & (0.0007) \\
Run 8--0 $\cdot$ Illiquidity &  & 0.0021** \\
 &  & (0.0010) \\
\(|\Delta q_{it}|\) & 0.4057*** & 0.4122*** \\
 & (0.0080) & (0.0080) \\
\midrule
State controls & Yes & Yes \\
Observations & 179,471 & 179,471 \\
Game clusters & 1,438 & 1,438 \\
\bottomrule
\end{tabular}
\begin{minipage}{0.92\textwidth}
\vspace{0.1cm}
\footnotesize
Notes: The dependent variable is directional underreaction, \(UR_{it}=\operatorname{sign}(\Delta q_{it})(\Delta q_{it}-\Delta p_{it})\). Positive values indicate that the Kalshi price moved too little in the direction implied by the benchmark probability. Salience and illiquidity are standardized. State controls include lagged market and benchmark probabilities, score margin, absolute score margin, game-clock minutes remaining, close-game and clutch indicators. The sample excludes negligible benchmark innovations, \(|\Delta q_{it}|<0.0025\). Standard errors, in parentheses, are clustered by game. ***, **, and * denote significance at the 1, 5, and 10 percent levels.
\end{minipage}
\end{table}

\begin{figure}[h!]
\centering
\includegraphics[width=0.78\textwidth]{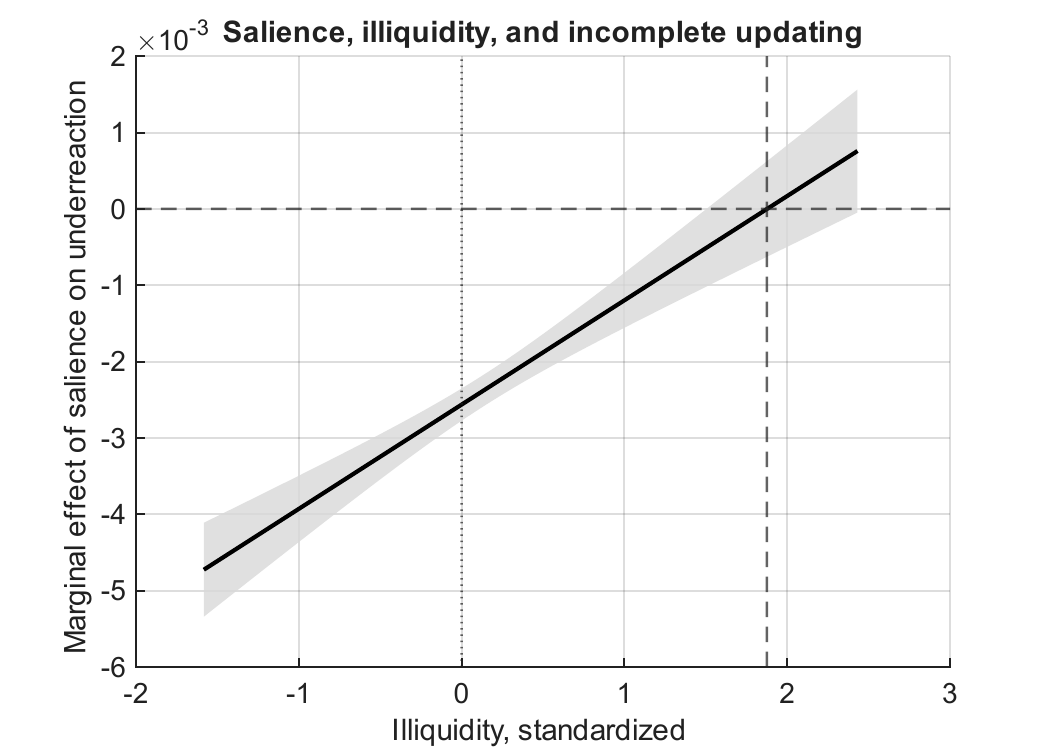}
\caption{Marginal effect of salience on incomplete updating by illiquidity}
\label{fig:marginal_salience_illiquidity}
\begin{minipage}{0.88\textwidth}
\vspace{0.1cm}
\footnotesize
Notes: The figure reports the marginal effect of salience on impact underreaction as a function of standardized illiquidity. Impact underreaction is defined as \(\operatorname{sign}(\Delta q_{it})(\Delta q_{it}-\Delta p_{it})\), so positive values indicate that the market moved too little in the direction implied by the benchmark. The shaded area denotes 95 percent confidence intervals based on game-clustered standard errors.
\end{minipage}
\end{figure}

Column 2 confirms this pattern using individual event indicators. Three-point shots, turnovers, lead changes, and large scoring runs are all associated with smaller underreaction gaps on average. These events attract attention and appear to facilitate immediate information processing. However, the interaction with illiquidity is positive for three-point shots, lead changes, and scoring runs. The effect is largest for lead changes. A lead change is among the most visible events in a game, yet in a thin market it is not fully incorporated on impact. This result is important because it rejects a simple salience-as-bias interpretation. In our data, salient information is not systematically overreacted to. Instead, salience appears to improve immediate recognition of information, while liquidity determines whether that information is completely reflected in prices. Figure~\ref{fig:marginal_salience_illiquidity} illustrates the interaction. It plots the marginal effect of salience on directional underreaction,
\begin{equation}
\label{eq:marginal_salience}
\frac{\partial UR_{it}}{\partial Salience_{it}}
=
\beta+\theta Illiquidity_{it}.
\end{equation}
The marginal effect is negative when markets are relatively liquid, implying that salient signals are incorporated more completely. As illiquidity increases, the effect becomes less negative and eventually turns positive. The same public signal can therefore be incorporated rapidly in a liquid market but only partially in a thin market.

We next examine whether the gradual adjustment documented in Section~\ref{sec:live} also depends on salience and liquidity. To do so, we augment Equation~\eqref{eq:drift1} with interactions between the updating gap, salience, and illiquidity.
Specifically, we estimate
\[
\left(p_{i,t+h}-p_{it}\right)
-
\left(q_{i,t+h}-q_{it}\right)
=
\alpha
+
\rho Gap_{it}
+
\phi Gap_{it}\cdot Illiquidity_{it}
+
\psi Gap_{it}\cdot Salience_{it}
+
\Gamma X_{it}
+
\varepsilon_{i,t+h}.
\]
The dependent variable is future price drift net of future benchmark changes. The coefficient \(\rho\) captures the average correction of the initial updating gap. The interaction terms test whether this correction differs across salient and illiquid states. Table~\ref{tab:mechanisms_drift} reports the results for horizons of 5, 10, and 15 minutes. The updating gap strongly predicts subsequent drift at all horizons, confirming the gradual-adjustment result from the previous section. More importantly, the interaction with salience is positive and statistically significant throughout. Underreaction gaps associated with more salient states are followed by larger subsequent corrections. Salient information that is not fully incorporated immediately continues to enter prices over the following minutes. The interaction with illiquidity is not positive in the drift regressions. If anything, the coefficient becomes negative at longer horizons. This suggests that illiquidity increases the initial failure to incorporate information but does not accelerate subsequent correction. Instead, the same trading frictions that slow adjustment on impact may also slow convergence afterward. Taken together, the evidence points to gradual price discovery shaped by the interaction between attention and trading frictions. Salient public signals are not ignored: in liquid markets they are incorporated relatively quickly. However, when salient information arrives in thin markets, prices adjust less completely on impact and continue to drift in the benchmark direction over subsequent minutes. The mechanism is therefore neither simple inattention nor systematic overreaction. Rather, real-time markets appear to recognize salient public information, while liquidity determines how fully that information is incorporated into prices. This interaction between behavioral attention and market microstructure helps explain why even simple binary contracts with transparent public signals can exhibit predictable short-horizon drift.

\begin{table}[h!]\centering
\caption{Salience, liquidity, and future price drift}
\label{tab:mechanisms_drift}
\begin{tabular}{lccc}
\toprule
 & (1) & (2) & (3) \\
 & 5 min & 10 min & 15 min \\
\midrule
Updating gap & 0.502*** & 0.514*** & 0.549*** \\
 & (0.015) & (0.016) & (0.016) \\
Gap $\cdot$ Illiquidity & -0.022 & -0.032 & -0.037** \\
 & (0.017) & (0.020) & (0.015) \\
Gap $\cdot$ Salience & 0.023*** & 0.013** & 0.025*** \\
 & (0.006) & (0.006) & (0.006) \\
\midrule
Controls & Yes & Yes & Yes \\
Observations & 171,790 & 164,350 & 156,670 \\
Game clusters & 1,438 & 1,438 & 1,437 \\
\bottomrule
\end{tabular}
\begin{minipage}{0.92\textwidth}
\vspace{0.1cm}
\footnotesize
Notes: The dependent variable is future price drift net of future benchmark changes, \((p_{i,t+h}-p_{it})-(q_{i,t+h}-q_{it})\). The underreaction gap is \(\Delta q_{it}-\Delta p_{it}\). Salience and illiquidity are standardized. Controls include \(\Delta q_{it}\), lagged market and benchmark probabilities, score margin, absolute score margin, game-clock minutes remaining, close-game and clutch indicators. Standard errors, in parentheses, are clustered by game. ***, **, and * denote significance at the 1, 5, and 10 percent levels, respectively.
\end{minipage}
\end{table}

\section{Conclusions}
\label{sec:conclusion}

Prediction markets are increasingly used as real-time measures of collective beliefs. Their usefulness depends not only on whether prices are calibrated on average, but also on whether they update efficiently when public information arrives. This paper studies that question using live NBA event contracts traded on Kalshi and merged with high-frequency play-by-play data. The setting allows us to observe market prices, liquidity, public signals, and terminal payoffs at high frequency, and to compare price changes with changes in a public-information benchmark win probability. The evidence supports a nuanced view of prediction market efficiency. Kalshi prices are highly informative. Pre-game prices are well calibrated, become more accurate as game time approaches, and provide a meaningful prior for the live market. During games, prices respond rapidly to public information: favorable scoring, made shots, opponent turnovers, lead changes, and scoring runs all move prices in the expected direction. In this sense, the market processes public information in real time. However, directional responsiveness is not the same as efficient updating. Relative to an out-of-sample benchmark win-probability model, live prices underreact on impact. A one-minute change in benchmark win probability is associated with only about a 0.64-for-one contemporaneous change in the Kalshi midpoint. The missing adjustment predicts future price drift over the next several minutes, including drift net of subsequent changes in the benchmark probability. Prices therefore move in the right direction, but not far enough. The mechanism analysis shows that incomplete updating is shaped by both attention and trading frictions. Salient events are incorporated relatively quickly in liquid markets, suggesting that visible public signals attract attention and are processed rapidly. But when salient information arrives in thin markets, prices adjust less completely on impact and continue to drift in the benchmark direction. Salience alone does not imply bias, and liquidity alone does not determine efficiency. The failure arises from their interaction: markets appear to recognize salient public information, while liquidity determines how fully that information is incorporated into prices. These findings have implications for how prediction market prices should be interpreted. Prediction markets can provide useful and timely forecasts, and our evidence confirms that their prices contain substantial information. But informativeness is not instantaneous efficiency. When public information arrives rapidly, especially in low-liquidity states, market-implied probabilities may be directionally correct yet temporarily incomplete. This distinction matters for researchers, firms, and policymakers who use prediction market prices as measures of beliefs. More broadly, the paper shows that live prediction markets are valuable laboratories for studying belief updating. They combine binary payoffs, precisely timed public signals, real money, and endogenous liquidity. This environment makes it possible to study not only whether markets aggregate information, but when and why the aggregation process is incomplete. The evidence suggests that real-time markets process public information quickly, but not always fully: price discovery is gradual and predictably shaped by the interaction of attention and trading frictions.

\newpage
\bibliographystyle{apalike}
\bibliography{references}

@article{AmihudMendelson1986,
  author  = {Amihud, Yakov and Mendelson, Haim},
  title   = {Asset Pricing and the Bid--Ask Spread},
  journal = {Journal of Financial Economics},
  year    = {1986},
  volume  = {17},
  number  = {2},
  pages   = {223--249},
  doi     = {10.1016/0304-405X(86)90065-6}
}

@article{AngeliniDeAngelisSingleton2022,
  author  = {Angelini, Giovanni and De Angelis, Luca and Singleton, Carl},
  title   = {Informational Efficiency and Behaviour within In-Play Prediction Markets},
  journal = {International Journal of Forecasting},
  year    = {2022},
  volume  = {38},
  number  = {1},
  pages   = {282--299},
  doi     = {10.1016/j.ijforecast.2021.05.012}
}

@article{BarberisShleiferVishny1998,
  author  = {Barberis, Nicholas and Shleifer, Andrei and Vishny, Robert},
  title   = {A Model of Investor Sentiment},
  journal = {Journal of Financial Economics},
  year    = {1998},
  volume  = {49},
  number  = {3},
  pages   = {307--343},
  doi     = {10.1016/S0304-405X(98)00027-0}
}

@article{BordaloGennaioliShleifer2012,
  author  = {Bordalo, Pedro and Gennaioli, Nicola and Shleifer, Andrei},
  title   = {Salience Theory of Choice Under Risk},
  journal = {Quarterly Journal of Economics},
  year    = {2012},
  volume  = {127},
  number  = {3},
  pages   = {1243--1285},
  doi     = {10.1093/qje/qjs018}
}

@techreport{BurgiDengWhelan2026,
  author      = {B{\"u}rgi, Constantin and Deng, Wanying and Whelan, Karl},
  title       = {Makers and Takers: The Economics of the Kalshi Prediction Market},
  institution = {CESifo},
  type        = {Working Paper},
  number      = {12122},
  year        = {2026}
}

@article{CroxsonReade2014,
  author  = {Croxson, Karen and Reade, J. James},
  title   = {Information and Efficiency: Goal Arrival in Soccer Betting},
  journal = {Economic Journal},
  year    = {2014},
  volume  = {124},
  number  = {575},
  pages   = {62--91},
  doi     = {10.1111/ecoj.12033}
}

@article{DanielHirshleiferSubrahmanyam1998,
  author  = {Daniel, Kent and Hirshleifer, David and Subrahmanyam, Avanidhar},
  title   = {Investor Psychology and Security Market Under- and Overreactions},
  journal = {Journal of Finance},
  year    = {1998},
  volume  = {53},
  number  = {6},
  pages   = {1839--1885},
  doi     = {10.1111/0022-1082.00077}
}

@article{DellaVignaPollet2009,
  author  = {DellaVigna, Stefano and Pollet, Joshua M.},
  title   = {Investor Inattention and Friday Earnings Announcements},
  journal = {Journal of Finance},
  year    = {2009},
  volume  = {64},
  number  = {2},
  pages   = {709--749}
}

@article{Fama1970,
  author  = {Fama, Eugene F.},
  title   = {Efficient Capital Markets: A Review of Theory and Empirical Work},
  journal = {Journal of Finance},
  year    = {1970},
  volume  = {25},
  number  = {2},
  pages   = {383--417},
  doi     = {10.2307/2325486}
}

@article{GauriotPage2026,
  author  = {Gauriot, Romain and Page, Lionel},
  title   = {How Market Prices React to Information: Evidence from Binary Options Markets},
  journal = {Economic Journal},
  year    = {2026},
  volume  = {136},
  number  = {673},
  pages   = {163--183},
  doi     = {10.1093/ej/ueaf040}
}

@article{GauriotPage2018,
  author  = {Gauriot, Romain and Page, Lionel},
  title   = {Psychological Momentum in Contests: The Case of Scoring Before Half-Time in Football},
  journal = {Journal of Economic Behavior \& Organization},
  year    = {2018},
  volume  = {149},
  pages   = {137--168},
  doi     = {10.1016/j.jebo.2018.02.015}
}

@article{GjerstadHall2005,
  author  = {Gjerstad, Steven and Hall, Michael C.},
  title   = {Risk Aversion, Beliefs, and Prediction Market Equilibrium},
  journal = {Economic Science Laboratory Working Paper, University of Arizona},
  year    = {2005}
}

@article{GlostenMilgrom1985,
  author  = {Glosten, Lawrence R. and Milgrom, Paul R.},
  title   = {Bid, Ask and Transaction Prices in a Specialist Market with Heterogeneously Informed Traders},
  journal = {Journal of Financial Economics},
  year    = {1985},
  volume  = {14},
  number  = {1},
  pages   = {71--100},
  doi     = {10.1016/0304-405X(85)90044-3}
}

@article{GrossmanStiglitz1980,
  author  = {Grossman, Sanford J. and Stiglitz, Joseph E.},
  title   = {On the Impossibility of Informationally Efficient Markets},
  journal = {American Economic Review},
  year    = {1980},
  volume  = {70},
  number  = {3},
  pages   = {393--408}
}

@article{HirshleiferLimTeoh2009,
  author  = {Hirshleifer, David and Lim, Sonya Seongyeon and Teoh, Siew Hong},
  title   = {Driven to Distraction: Extraneous Events and Underreaction to Earnings News},
  journal = {Journal of Finance},
  year    = {2009},
  volume  = {64},
  number  = {5},
  pages   = {2289--2325},
  doi     = {10.1111/j.1540-6261.2009.01501.x}
}

@article{HongStein1999,
  author  = {Hong, Harrison and Stein, Jeremy C.},
  title   = {A Unified Theory of Underreaction, Momentum Trading, and Overreaction in Asset Markets},
  journal = {Journal of Finance},
  year    = {1999},
  volume  = {54},
  number  = {6},
  pages   = {2143--2184},
  doi     = {10.1111/0022-1082.00184}
}

@article{Kyle1985,
  author  = {Kyle, Albert S.},
  title   = {Continuous Auctions and Insider Trading},
  journal = {Econometrica},
  year    = {1985},
  volume  = {53},
  number  = {6},
  pages   = {1315--1335},
  doi     = {10.2307/1913210}
}

@article{Manski2006,
  author  = {Manski, Charles F.},
  title   = {Interpreting the Predictions of Prediction Markets},
  journal = {Economics Letters},
  year    = {2006},
  volume  = {91},
  number  = {3},
  pages   = {425--429},
  doi     = {10.1016/j.econlet.2006.01.004}
}

@article{OettingDeutscherSingletonDeAngelis2025,
  author  = {{\"O}tting, Marius and Deutscher, Christian and Singleton, Carl and De Angelis, Luca},
  title   = {Betting on Momentum in Contests},
  journal = {Economic Inquiry},
  year    = {2025},
  volume  = {63},
  number  = {4},
  pages   = {1066--1089},
  doi     = {10.1111/ecin.70008}
}

@article{Page2012,
  author  = {Page, Lionel},
  title   = {{``}It Ain't Over Till It's Over.{''} {Yogi Berra} Bias on Prediction Markets},
  journal = {Applied Economics},
  year    = {2012},
  volume  = {44},
  number  = {1},
  pages   = {81--92}
}

@article{PageClemen2013,
  author  = {Page, Lionel and Clemen, Robert T.},
  title   = {Do Prediction Markets Produce Well-Calibrated Probability Forecasts?},
  journal = {Economic Journal},
  year    = {2013},
  volume  = {123},
  number  = {568},
  pages   = {491--513},
  doi     = {10.1111/j.1468-0297.2012.02561.x}
}

@article{Sauer1998,
  author  = {Sauer, Raymond D.},
  title   = {The Economics of Wagering Markets},
  journal = {Journal of Economic Literature},
  year    = {1998},
  volume  = {36},
  number  = {4},
  pages   = {2021--2064}
}

@article{ThalerZiemba1988,
  author  = {Thaler, Richard H. and Ziemba, William T.},
  title   = {Anomalies: Parimutuel Betting Markets: Racetracks and Lotteries},
  journal = {Journal of Economic Perspectives},
  year    = {1988},
  volume  = {2},
  number  = {2},
  pages   = {161--174},
  doi     = {10.1257/jep.2.2.161}
}

@article{ShleiferVishny1997,
  author = {Shleifer, Andrei and Vishny, Robert W.},
  title = {The Limits of Arbitrage},
  journal = {Journal of Finance},
  year = {1997},
  volume = {52},
  number = {1},
  pages = {35--55},
  doi = {10.1111/j.1540-6261.1997.tb03807.x}
}

@article{WolfersZitzewitz2004,
  author  = {Wolfers, Justin and Zitzewitz, Eric},
  title   = {Prediction Markets},
  journal = {Journal of Economic Perspectives},
  year    = {2004},
  volume  = {18},
  number  = {2},
  pages   = {107--126},
  doi     = {10.1257/0895330041371321}
}

@techreport{WolfersZitzewitz2006,
  author      = {Wolfers, Justin and Zitzewitz, Eric},
  title       = {Interpreting Prediction Market Prices as Probabilities},
  institution = {National Bureau of Economic Research},
  type        = {NBER Working Paper},
  number      = {12200},
  year        = {2006},
  doi         = {10.3386/w12200}
}

\clearpage
\appendix
\renewcommand{\thesection}{Appendix \Alph{section}}
\renewcommand{\thesubsection}{\Alph{section}.\arabic{subsection}}

\section{Alternative Benchmark Models}
\label{app:benchmark_models}

A natural concern is that the incomplete-updating result reflects misspecification of the benchmark win-probability model rather than incomplete incorporation of public information. To address this concern, we re-estimate the main live-updating tests using several alternative benchmark specifications. All benchmarks predict the final contract payoff using only pre-game information and public game-state variables and exclude contemporaneous Kalshi live prices. We consider five specifications. The baseline model is the benchmark used in the main text. The parsimonious specification uses a smaller set of game-state controls. The no-recent-scoring specification excludes recent net-scoring variables. The flexible specification adds higher-order terms and additional nonlinear state interactions. Finally, the chronological-holdout specification estimates the model on the first 70 percent of games and evaluates it on the remaining 30 percent. For the first four specifications, benchmark probabilities are estimated using five-fold cross-fitting at the game level. We then compute
\[
\Delta q_{it}=q_{it}-q_{i,t-1}
\]
and re-estimate both the contemporaneous updating regression,
\[
\Delta p_{it}
=
\alpha
+
\beta \Delta q_{it}
+
\Gamma X_{it}
+
\varepsilon_{it},
\]
and the five-minute drift specification,
\[
\left(p_{i,t+5}-p_{it}\right)
-
\left(q_{i,t+5}-q_{it}\right)
=
\alpha
+
\rho\left(\Delta q_{it}-\Delta p_{it}\right)
+
\Gamma X_{it}
+
\varepsilon_{i,t+5}.
\]

Table~\ref{tab:robust_benchmark_models} reports the results. Across all benchmark specifications, the coefficient on \(\Delta q_{it}\) remains well below one, ranging from 0.637 to 0.698. In every case, the null of one-for-one updating is strongly rejected. The underreaction gap also continues to predict future price drift net of future benchmark changes. The corresponding coefficients range from 0.379 to 0.493 and remain statistically significant throughout. These results indicate that the main finding is not driven by a particular specification of the benchmark win-probability model. Live prices respond strongly to public-information updates, but the contemporaneous response remains incomplete across parsimonious, flexible, no-recent-scoring, and chronological-holdout benchmarks.

\begin{table}[h!]\centering
\caption{Robustness to alternative benchmark win-probability models}
\label{tab:robust_benchmark_models}
\begin{tabular}{lccc}
\toprule
Benchmark model & Brier \(q\) & \(\beta\) on \(\Delta q\) & 5-min net drift \\
\midrule
Baseline logit & 0.164 & 0.638 & 0.459*** \\
 &  & (0.010) & (0.029) \\
Parsimonious logit & 0.164 & 0.676 & 0.379*** \\
 &  & (0.007) & (0.020) \\
No recent scoring & 0.164 & 0.640 & 0.460*** \\
 &  & (0.010) & (0.030) \\
Flexible state logit & 0.164 & 0.637 & 0.460*** \\
 &  & (0.010) & (0.030) \\
Chronological split & 0.138 & 0.698 & 0.493*** \\
 &  & (0.010) & (0.028) \\
\bottomrule
\end{tabular}
\begin{minipage}{0.92\textwidth}
\vspace{0.1cm}
\footnotesize
Notes: The table reports robustness to alternative benchmark win-probability models. The coefficient on \(\Delta q\) comes from regressions of one-minute Kalshi midpoint changes on one-minute benchmark probability changes and controls. Controls include the lagged midpoint, lagged benchmark probability, score margin, absolute score margin, game-clock minutes remaining, bid--ask spread, recent volume, and open interest. The five-minute net drift coefficient is the coefficient on the underreaction gap, \(\Delta q_{it}-\Delta p_{it}\), in regressions where the dependent variable is \((p_{i,t+5}-p_{it})-(q_{i,t+5}-q_{it})\). The chronological split is evaluated only on the holdout period; its Brier score and sample size are therefore not directly comparable to the cross-fitted specifications. Standard errors, in parentheses, are clustered by game. ***, **, and * denote significance at the 1, 5, and 10 percent levels, respectively.
\end{minipage}
\end{table}

\section{Microstructure and Quote Quality}
\label{app:microstructure}

A second concern is that the incomplete-updating result reflects stale quotes or other microstructure effects rather than gradual incorporation of public information. We therefore re-estimate the main updating and drift regressions on a series of quote-quality subsamples. Specifically, we consider observations with positive trading volume in the current minute, positive trading volume over the previous five minutes, bid--ask spreads below the sample median, and samples excluding stale no-trade quotes. We also construct a high-quality quote sample combining narrow spreads, positive recent volume, and non-stale quotes. For each sample, we re-estimate
\[
\Delta p_{it}
=
\alpha
+
\beta \Delta q_{it}
+
\Gamma X_{it}
+
\varepsilon_{it},
\]
and
\[
\left(p_{i,t+5}-p_{it}\right)
-
\left(q_{i,t+5}-q_{it}\right)
=
\alpha
+
\rho\left(\Delta q_{it}-\Delta p_{it}\right)
+
\Gamma X_{it}
+
\varepsilon_{i,t+5}.
\]

Table~\ref{tab:robust_microstructure} reports the results. The coefficient on \(\Delta q_{it}\) remains between 0.656 and 0.663 across quote-quality subsamples, while the five-minute drift coefficient remains positive and highly significant throughout. Thus, the incomplete contemporaneous response is not driven by inactive periods, wide spreads, or stale midpoint quotes.

\begin{table}[h!]\centering
\caption{Microstructure robustness}
\label{tab:robust_microstructure}
\begin{tabular}{lccc}
\toprule
Sample & \(\beta\) on \(\Delta q\) & 5-min net drift & Observations \\
\midrule
All observations & 0.638 & 0.459*** & 356,769 \\
 & (0.010) & (0.029) & \\
Positive volume, current minute & 0.657 & 0.533*** & 350,560 \\
 & (0.005) & (0.009) & \\
Positive volume, last 5 minutes & 0.656 & 0.531*** & 353,240 \\
 & (0.005) & (0.009) & \\
Narrow spread & 0.661 & 0.494*** & 257,200 \\
 & (0.005) & (0.015) & \\
Exclude stale/no-trade quotes & 0.657 & 0.533*** & 351,880 \\
 & (0.005) & (0.009) & \\
High-quality quotes & 0.663 & 0.507*** & 256,060 \\
 & (0.004) & (0.010) & \\
\bottomrule
\end{tabular}
\begin{minipage}{0.92\textwidth}
\vspace{0.1cm}
\footnotesize
Notes: The table reports robustness to microstructure-related sample restrictions. The coefficient on \(\Delta q\) comes from regressions of one-minute Kalshi midpoint changes on one-minute benchmark probability changes and controls. The five-minute net drift coefficient is the coefficient on the updating gap, \(\Delta q_{it}-\Delta p_{it}\), in regressions where the dependent variable is \((p_{i,t+5}-p_{it})-(q_{i,t+5}-q_{it})\). Controls include the lagged midpoint, lagged benchmark probability, score margin, absolute score margin, game-clock minutes remaining, bid--ask spread, recent volume, and open interest. Standard errors are clustered by game. ***, **, and * denote significance at the 1, 5, and 10 percent levels, respectively.
\end{minipage}
\end{table}

We next examine whether the predictable midpoint drift documented in the main text corresponds to an executable trading opportunity. Define the underreaction gap as
\[
Gap_{it}
=
\Delta q_{it}-\Delta p_{it}.
\]
Positive gaps generate long positions in the YES contract, while negative gaps generate short positions. We compare midpoint returns with executable-style returns that explicitly account for bid--ask spreads. Table~\ref{tab:executable_returns} reports the results. Midpoint returns are positive and increase with the size of the underreaction gap, confirming that incomplete updating predicts subsequent midpoint movements in the expected direction. However, executable-style returns remain negative even for large gaps. Once transaction costs are incorporated, the apparent predictability is largely absorbed by the bid--ask spread.

\begin{table}[h!]\centering
\caption{Directional returns after underreaction gaps}
\label{tab:executable_returns}
\begin{tabular}{llccc}
\toprule
Gap threshold & Sample & Midpoint return & Executable-style return & Observations \\
\midrule
\(|Gap|\geq 0\) bp & All & 0.0039*** & -0.0120*** & 300,920 \\
 & High-quality & 0.0037*** & -0.0078*** & 212,780 \\
\addlinespace
\(|Gap|\geq 50\) bp & All & 0.0050*** & -0.0109*** & 219,990 \\
 & High-quality & 0.0049*** & -0.0067*** & 150,780 \\
\addlinespace
\(|Gap|\geq 100\) bp & All & 0.0062*** & -0.0101*** & 163,490 \\
 & High-quality & 0.0059*** & -0.0057*** & 112,420 \\
\addlinespace
\(|Gap|\geq 200\) bp & All & 0.0087*** & -0.0092*** & 92,665 \\
 & High-quality & 0.0082*** & -0.0036*** & 60,965 \\
\bottomrule
\end{tabular}
\begin{minipage}{0.92\textwidth}
\vspace{0.1cm}
\footnotesize
Notes: The table reports five-minute directional returns following underreaction gaps. If \(Gap_{it}=\Delta q_{it}-\Delta p_{it}>0\), the strategy buys the YES contract; if \(Gap_{it}<0\), it sells the YES contract. The midpoint return uses midpoint prices. The executable-style return buys at the ask and exits at the bid for long positions, and sells at the bid and covers at the ask for short positions. When actual bid and ask quotes are unavailable, bid and ask are approximated as midpoint minus and plus half the quoted spread. Standard errors are clustered by game. ***, **, and * denote significance at the 1, 5, and 10 percent levels, respectively.
\end{minipage}
\end{table}

Taken together, the robustness exercises support two conclusions. First, the main incomplete-updating result is not sensitive to the specification of the benchmark model and is not driven by stale quotes or low-quality observations. Second, the predictable drift documented in midpoint prices should not be interpreted as a frictionless arbitrage opportunity. The evidence is best interpreted as gradual price discovery in market-implied probabilities under trading frictions.

\section{Heterogeneity by Game Phase}
\label{app:game_phase}

A further concern is that the incomplete-updating result may be concentrated in a narrow part of the game. For example, late-game states are more volatile, benchmark win probabilities change more sharply, and market liquidity may differ from earlier phases. If the coefficient below one were driven only by these states, the main result would be less general. We therefore re-estimate the live-updating regression separately by game phase.

For each subsample, we estimate

\[
\Delta p_{it}
=
\alpha
+
\beta \Delta q_{it}
+
\Gamma X_{it}
+
\varepsilon_{it},
\]

where \(X_{it}\) includes the same controls as in the baseline specification. Standard errors are clustered by game. We consider the first, second, third, and fourth quarters separately. We also split fourth-quarter observations into non-clutch and clutch states. Clutch time is defined as the final five minutes of the fourth quarter with an absolute score margin of five points or less. Table~\ref{tab:robust_game_phase} reports the results. The coefficient on benchmark probability changes remains well below one in every phase of the game. The baseline coefficient is 0.638. Across the first three quarters, the estimates range from 0.641 to 0.673. In the fourth quarter, the coefficient is 0.620. The non-clutch fourth-quarter coefficient is 0.700, while the clutch-time coefficient is lower, at 0.509. In all cases, the null of one-for-one updating is strongly rejected. These results show that incomplete updating is not driven by a narrow subset of late-game observations. Underreaction is present throughout the game. It is especially pronounced in clutch states, where public-information shocks are likely to be more salient, win probabilities are more sensitive to small changes in the game state, and trading frictions may matter more. This pattern reinforces the interpretation that real-time prices respond to public information but incorporate it only gradually.

\begin{table}[h!]\centering
\caption{Live updating by game phase}
\label{tab:robust_game_phase}
\begin{tabular}{lccccc}
\toprule
Sample & Observations & Games & \(\beta\) on \(\Delta q\) & SE & \(p(\beta=1)\) \\
\midrule
All observations & 356,769 & 1,438 & 0.638 & 0.010 & $<0.001$ \\
First quarter & 78,811 & 1,438 & 0.641 & 0.005 & $<0.001$ \\
Second quarter & 123,690 & 1,435 & 0.663 & 0.005 & $<0.001$ \\
Third quarter & 87,568 & 1,410 & 0.673 & 0.006 & $<0.001$ \\
Fourth quarter & 66,700 & 1,244 & 0.620 & 0.009 & $<0.001$ \\
Non-clutch fourth quarter & 51,844 & 1,244 & 0.700 & 0.008 & $<0.001$ \\
Clutch time & 14,856 & 616 & 0.509 & 0.012 & $<0.001$ \\
\bottomrule
\end{tabular}
\begin{minipage}{0.92\textwidth}
\vspace{0.1cm}
\footnotesize
Notes: The table reports regressions of one-minute Kalshi midpoint changes on one-minute benchmark probability changes, estimated separately by game phase. The dependent variable is \(\Delta p_{it}\). The main regressor is \(\Delta q_{it}\), the one-minute change in the benchmark win probability. Controls are the same as in the baseline live-updating specification. Standard errors are clustered by game. Clutch time is defined as the final five minutes of the fourth quarter with an absolute score margin of five points or less. The last column reports the \(p\)-value for the test \(H_0:\beta=1\).
\end{minipage}
\end{table}

\end{document}